# An algorithm for Bragg coherent X-ray diffractive imaging of highly strained nanocrystals

*Ziyi Wang, Oleg Gorobtsov, Andrej Singer*

*Department of Materials Science and Engineering, Cornell University*

**Abstract**

By using phase retrieval, Bragg Coherent Diffractive Imaging (BCDI) allows tracking of three-dimensional displacement fields inside individual nanocrystals. Nevertheless, in the presence of significant (1% and higher) strains, such as in the process of a structural phase transformation, fails due to the Bragg peak distortions. Here we present an advanced BCDI algorithm enabling imaging three-dimensional strain fields in highly strained crystals. We test the algorithm on particles simulated to undergo a structural phase transformation. While the conventional algorithm fails in unambiguously reconstructing the phase morphology, our algorithm correctly retrieves the morphology of coexistent phases with a strain difference of 1%. The key novelty is the simultaneous reconstruction of multiple scans of the same nanoparticle at snapshots through the phase transformations. The algorithm enables visualizing phase transformations in nanoparticles of lithium-ion, sodium-ion nanoparticles, and other nanoparticulate materials in working conditions (operando).

**Introduction**

Thermodynamics and kinetics of phase transformations play a vital role in materials science and physics. The solid-solid structural rearrangements – specifically the phase morphology at the sub-micron length scale – dictate the properties of structural materials[1], quantum materials[2], and energy storage materials [3–6]. These materials often consist of micron-sized crystalline particles or grains; accessing the nanoscale morphology inside these particles during phase transformation presents a critical scientific challenge necessitating *in-situ* imaging techniques. A battery of powerful tools exists for analyzing the phase transformations in materials. X-ray diffraction (XRD)[7,8] characterizes the crystal structure; nevertheless, traditional powder XRD only accesses the average properties of a large ensemble of particles and yields the average phase fractions[8,9]. Isolating a specific nanoparticle through conventional XRD is impossible. Microscopy methods, including electron diffraction[9] and electron microscopy[9,10], are still challenging in functional devices. Therefore, an *in-situ* tool for imaging the real-time lattice response inside single nanoparticles with high resolution likely will offer new insight.

With the development of partially coherent synchrotron X-ray sources, Bragg coherent diffractive imaging[11] (BCDI) emerged as a tool for imaging three-dimensional (3D) strain fields inside crystalline nanoparticles.[12,13] BCDI consists of two processes: (1) recording a 3D reciprocal space intensity around the Bragg peak with coherent X-rays and (2) subsequently phasing this intensity with phase retrieval algorithms for reconstructing a real space image. BCDI enabled studying strain and defect distributions in single free-standing nanoparticles.[14,15] BCDI is also capable of the *in-situ* detection of strain evolution in nanocrystals undergoing chemical reactions[16,17], or battery nanoparticles undergoing charge-discharge cycles[18–21]. BCDI is an ideal tool for studying phase transformations, as the coexisting structural phases have a different crystal structure resulting in readily detectable displacement fields. However, limitations of the current BCDI method become apparent when the magnitude of the strain field is high, and the Bragg peak significantly distorts or even splits. For nanoparticles with (500nm)$^3$ size, the magnitude of these strain fields is typically around 1%.[8,10] In these cases, the previous attempts to reconstruct the nanoparticles have been unsuccessful[22].

We developed a simulation package for accessing the applicability of the current phase retrieval algorithms used with BCDI for imaging phase transformations. We simulated nanocrystals with coexisting phases in different morphologies, generated noisy diffraction data from these nanocrystals, applied phase retrieval algorithms, and compared the retrieved images with the input. We find that the current phase retrieval algorithm (used in [19,21]) fails when the lattice mismatch between the two phases is on the order of 1%, independent of the phase morphology. We developed a new algorithm and tested it on simulated data; the algorithm correctly retrieves the phase morphologies with a lattice mismatch of ~1 % in single nanoparticles. This paper illustrates the simulation package, outlines the new algorithm, and describes the performance tests we conducted on simulated noisy X-ray data.

Figure 1 shows a typical BCDI experimental setup. As a concrete example for our simulations, we consider an operando experiment similar to our previous experiments[18–21]. A coherent synchrotron X-ray beam focused to about 1 μm illuminates a nanoparticle, potentially embedded in a multicomponent operational device. A 2D detector collects the diffraction data at the Bragg angle. Through stepwise rocking of the sample by about $\pm 1°$, one records a series of 2D sections of the Ewald sphere around the Bragg peak[11]. These approximately parallel sections constitute the 3D diffraction intensity $I(\mathbf{Q})$. Consider a particle, which undergoes a structural phase transformation (for example, during the battery operation). The two coexisting phases have different lattice parameters $d_h + \Delta d$ and $d_h - \Delta d$ resulting in the splitting of the Bragg peak (see Figure 1 a, b). While the relative fractions are evident directly from the diffraction data (see Figure 1 c), quantitative imaging of the phase morphology requires inverting the measured intensity distribution through phase retrieval algorithms.

The 3D coherent X-ray scattering amplitude $A(\mathbf{Q})$ from a nanocrystal is well approximated as[23,24]

$$A(\mathbf{Q}) = F_h \int s(\mathbf{r}) e^{-i\mathbf{h}\cdot\mathbf{u}(\mathbf{r})} e^{-i\mathbf{Q}\cdot\mathbf{r}} d\mathbf{r}, \tag{1}$$

where $\mathbf{Q} = \mathbf{q} - \mathbf{h}$, $\mathbf{q} = \mathbf{k}_s - \mathbf{k}_i$ is the momentum transfer defined by the incident and scattered wave vectors, $\mathbf{h}$ is the reciprocal space vector, and $F_h$ is the structure factor of the measured Bragg peak. The measured intensity is $I(\mathbf{Q}) = |A(\mathbf{Q})|^2$. The real-space representation of the nanocrystal consists of the Ewald function (or shape function) $s(\mathbf{r})$ [23], with $s(\mathbf{r}) = 0$ outside and $s(\mathbf{r}) = 1$ inside of the nanocrystals, and the complex phase $\exp[-i\mathbf{h}\cdot\mathbf{u}(\mathbf{r})]$, where $\mathbf{u}(\mathbf{r})$ is the displacement of the atomic planes perpendicular to $\mathbf{h}$ and separated by $d_h = 2\pi/|\mathbf{h}|$. In short, the coherent scattering amplitude is the Fourier transform of $s(\mathbf{r}) \cdot \exp[-i\mathbf{h}\cdot\mathbf{u}(\mathbf{r})]$.

Inverting Equation (1) and thereby determining the 3D displacement field inside a nanocrystal requires finding the complex phases of $A(\mathbf{Q})$ lost in an intensity measurement. In the last decades, multiple phase retrieval algorithms emerged, including the Error-Reduction (ER) algorithm[25], the Hybrid-Input-Output (HIO) algorithm[26], Difference Map (DM)[27], and the Relaxed Averaged Alternating Reflection (RAAR) algorithm[28]. In the last two decades, these algorithms found successful applications in the X-ray community[29–32]. The algorithms alternate between the Fourier transform and its inverse (see Equation 1 or an equivalent equation for a different experimental geometry). Additionally, after the forward Fourier transform, the reconstructed intensities are replaced with the measured intensities $I(\mathbf{Q})$ while the complex phases remain. After the inverse Fourier transform, the real-space representation of the nanocrystal $s(\mathbf{r}) \cdot \exp[-i\mathbf{h}\cdot\mathbf{u}(\mathbf{r})]$ is forced into a finite volume by using the support $\sup(\mathbf{r})$, which is equal to 1 inside a volume and 0 outside (in the most straightforward ER algorithm, one multiplies the $s(\mathbf{r})$ by $\sup(\mathbf{r})$). Finding the correct support is vital for the convergence of the iterative algorithm because the shape of the crystal is generally unknown. The shrink-wrap algorithm[33] dynamically modifies the support by adapting it to the shape $s(r)$ and was a key discovery for the success of in-situ BCDI.

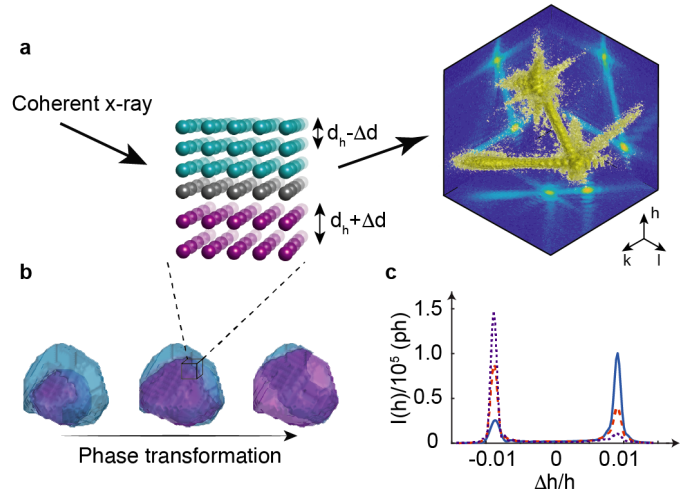

**Figure 1**. (a) A schematic of the BCDI experimental setup. (b) A particle is simulated to undergo a structural phase transformation from the small-unit-cell phase (blue) to the large-unit-cell phase (magenta). The simulated 3D diffraction pattern in (a) corresponds to the phase coexistence state. (c) The simulated powder XRD signal showing the phase fraction during the phase transformation from the small-unit-cell phase (dotted magenta line), through the phase coexistence (red dashed line), to the large-unit-cell phase (solid blue line).

**Methods**

In this work, we directly implement Equation 1. We first generate the shape $s(r)$ of the nanoparticle (see Equation 1) by constraining $s(r) = 1$ by 100 planes each with random orientation and a random distance (within a range $r_1$ and $r_2$) from the center of the coordinate system $r = 0$. After spatial smoothing by a Gaussian filter with a 7x7x7 pixels kernel, we reset the shape function to 0 and 1 with a threshold of 0.3. The smoothing and resetting remove the sharp features at the corners and edges of the nanoparticle. Repeating the above procedure three times generates an asymmetric nanoparticle shape (see Figure 1), common for battery nanoparticles[19–21].

For simulating the morphology of the two structural phases coexisting inside of the nanoparticle, we generate a surface inside the nanoparticle and set compressive strain $-\varepsilon_0$ and a tensile strain $+\varepsilon_0$, on different sides of the surface (see Figure 1b). This surface is the phase boundary between the coexisting structural phases. We smooth this interface by several pixels (a box filter with a 5x5x5 pixels kernel) to imitate the gradual change of the lattice constant due to coherency strain[34]. We calculate the strain $\pm\varepsilon_0 = (d(r) \pm d_h)/d_h$ with a fixed average lattice constant $d_h$ for the complete phase transformation process. The average lattice constant $d_h$ is the average between the lattice constants of the two participating structural phases, and $d(r)$ is the local lattice constant imaged with BCDI. For calculating the displacement field in Equation 1, we use the relation $\varepsilon_h(r) = du_h(r)/dr_h$[21], where $r_h$ is the real space coordinate parallel to $h$, $u_h(r)$ is the displacement, and $\varepsilon_h(r)$ is the strain resolved along $h$. We calculate a displacement field $u_h(r)$ by numerically integrating the strain field $\varepsilon_h(r)$ along $r_h$ (y-direction in the real space, see Figure 1b). We assume that the x-z plane at y=0 has no displacement, meaning no bending of the atomic planes. In a separate simulation, we confirmed that adding constant displacement on the x-z plane at y=0 makes no difference in the results.

With the shape and the displacement field of the nanoparticle defined, we calculate the diffraction intensity $|A(Q)|^2$ according to Equation 1 by using the fast Fourier transform. The size of the whole simulated real space is (128 pixels)³ corresponding to (2μm)³, the size of our simulated nanoparticles is around (500nm)³, the magnitude of strain $\varepsilon_0$ as $10^{-2}$, the average lattice constant is $d_h = 5$ Å, the X-ray photon energy is 9keV, the diffraction angle $2\theta$ is 15.85°, the sample-detector distance is 0.8m, and the angle step $\delta\theta$ is 0.004°. These parameters resemble real experimental conditions of operando BCDI imaging of battery nanoparticles[19–21]. After calculating the intensity, we set the number of photons detected in the full 3D reciprocal space to $10^6$, which is currently achievable at the third-generation synchrotron sources with transition metal oxide nanoparticles for battery cathodes[14]. We apply Poisson noise to the diffraction pattern. For minimizing the leakage effect, which occurs when the crystal truncation rods reach the edge of the 3D volume, we calculate the diffraction pattern in a (256 pixel)$^3$ volume and keep the central (128 pixel)$^3$ volume for further analysis. Figure 1a shows the noisy, simulated 3D Bragg peak from the nanoparticle shown in Figure 1b (middle).

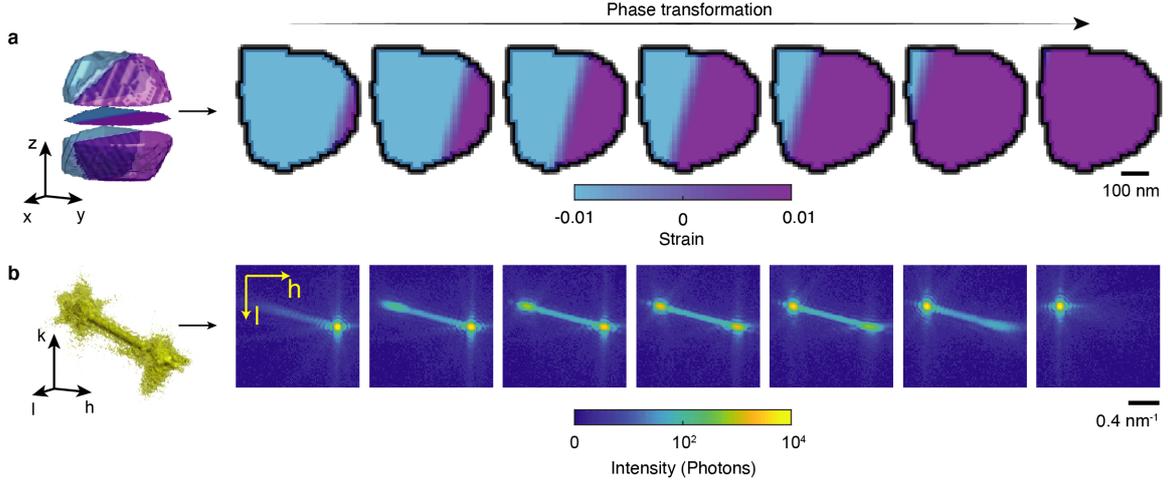

**Figure 2.** (a) The strain of a crystalline particle simulated during a structural phase transformation. The slice shown on the right is indicated in the 3D model on the left. (b) The corresponding simulated diffraction patterns in each state. The projection of the 3D diffraction pattern is shown for each state. The total number of photons in every 3D diffraction pattern is $10^6$.

A simulated nanocrystal undergoing a phase transformation and the corresponding 2D surfaces from the 3D reciprocal space are shown in Figure 2. In this example, the particle contains two structural phases in the left and right parts of the nanoparticle with a relatively flat phase boundary. Before the phase transformation, the particle consists wholly of a smaller unit cell (blue) phase. The new structural phase with a larger unit cell (purple) nucleates at the right boundary and grows gradually at the expense of the blue phase. We simulated the peak shape at seven points through the phase transformation in total. While for smaller strain ($\varepsilon_0 = 10^{-3}$) alternating HIO/ER or RAAR/ER algorithms lead to satisfactory results[35], applying the same algorithm on the datasets shown in Figure 2 with $\varepsilon_0 = 10^{-2}$ failed to retrieve the morphology of the phases correctly. The typical failures we observed included retrieving both structural phases in the same volume or having an incorrect spatial arrangement of the two phases (for instance, blue above purple instead of blue to the left of purple). Inverse Fourier transform of the noisy data combined with the known simulated complex phase of $A(\boldsymbol{Q})$ showed the correct morphology, indicating that the retrieving the phase morphology is in principle possible with a working phase retrieval algorithm.

Our algorithm introduces three additional procedures to the conventional phase retrieval (see Figure 3). First, and most importantly, we reconstruct multiple datasets shown in Figure 2 simultaneously; every few iterations, the algorithm pushes the reconstructions towards having a similar shape while leaving the displacement fields untouched (this approach resembles ptychography [31,36]). We think this assumption is applicable to most materials (for example, batteries), where the phase transformation occurs many times without modifying the performance necessitating negligible shape modifications. Specifically, after the real space constraint, we compute the average support from all reconstructions $\overline{\text{Sup}}(r) = \frac{1}{N} \cdot \sum_{\alpha=1} \sup_\alpha(r)$ (α=1 to N, N=7). After converting $\overline{\sup}(r)$ to values of 1 or 0 (threshold=0.6, ), we every individual support $\sup_\alpha(r)$ by $\overline{\sup}(r)$. This last step unifies the support from all different datasets with the same particle size and different morphology. We perform this operation after every tenth iteration.

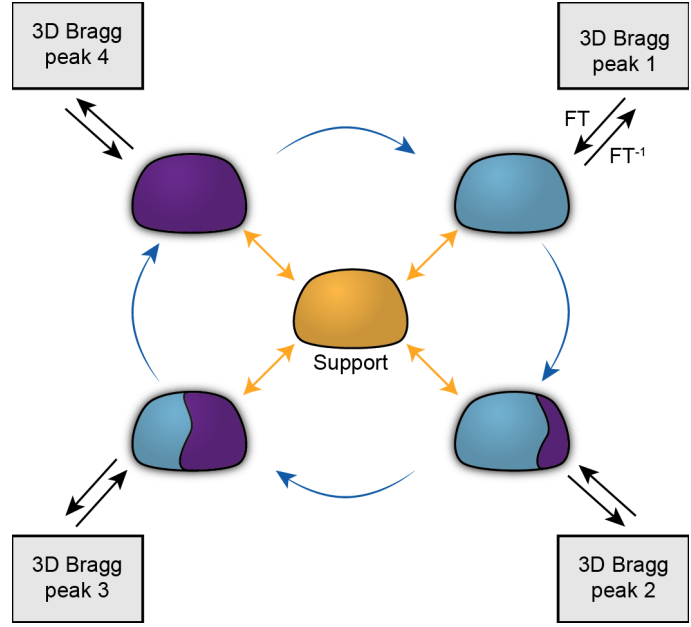

**Figure 3:** Schematic of our phase retrieval algorithm. An average support is generated by averaging all the supports of different scans, and this average is reset to 0 and 1 with a threshold of 0.6. This new support subsequently is multiplied onto the individual supports, unifying the reconstructions from separate measured 3D Bragg peaks.

Second, for every dataset shown in Figure 2 we simultaneously run 30 reconstructions, each initiated from a random set of initial phases. Out of those 30, we find the reconstruction most similar to all other reconstructions within a dataset by calculating the error matrix

$$M_{k,l} = E(\varepsilon_k, \varepsilon_l) + E(\sup_k, \overline{\sup})$$

where $E(g, h) = \sum(g-h)^2 / [\sum g^2 + \sum h^2]$ , and we sum over the whole volume in the real space. $\sup_k$ is the current support of the k-th reconstruction (out of 30), and $\overline{\sup}$ is the average support from the previous iteration found in the first procedure (from all scans shown in Figure 2). The first term identifies reconstructions (out of 30) with a similar strain field, while the second term penalizes large deviations of the support as compared with the previous support calculated from all seven scans shown in Figure 2. After calculating $M_k = \sum_l M_{k,l}$ and defining the best reconstruction to have the minimum $M_{k_m}$ , we find four other reconstructions with smallest $M_{k_m,l}$. The supports of these five best reconstructions are averaged to yield the current support for this dataset. Finally, all individual supports within a particular dataset are replaced with the corresponding support. We do the above procedure for every ten iterations, before calculating the average support for all the scans in step one. Note that we apply shrink-wrap[33] every five iterations (after 4th, 9th, 14th…). Noteworthy, in the first two steps we only modify the support and keep the shape function $s(r)$ and the complex phase $\mathbf{h}\cdot\mathbf{u}(r)$ untouched.

Third, after the two steps mentioned above and before the Fourier transform, we spatially smoothen the shape $s(r)$ with a box filter with a 3x3x3 pixels kernel. The smoothing effectively increases the extent of the support, preventing the two structural phases from being confined to an exceedingly compact region (the most common failure mechanism observed with the conventional algorithm). We apply smoothing in every iteration. Although the first two steps already present an improvement over the conventional algorithm, we found that the third step makes the phase retrieval algorithm significantly more robust. Finally, we start with 10 iterations of traditional ER algorithm, followed by 50 iterations of the RAAR algorithm. The first and second additional procedures are applied every ten iterations, 10(ER), 20 (RAAR), 30 (RAAR), …. In total, we use 610 iterations, while concluding with 10 iterations of ER.

**Results and discussion**

Figure 4 shows the reconstruction result of the nanoparticle shown in Figure 2 with the algorithm developed in this work. Compared with conventional phase retrieval algorithms, our new algorithm not only fully reconstructs the shape of the nanoparticle, but it also retrieves the morphology of the two structural phases with remarkably high reliability. The shape of the nanoparticle is well reconstructed and hardly changes between the different states. Both phases appear in the correct positions, and the magnitude of strain is consistent with the input used for calculating the diffraction patterns. Except for the insignificant inaccuracies in reconstructing the shape of the particles, the reconstructions are virtually identical to the initial data (compare Figure 4 with Figure 2a).

To verify the generality of our algorithm, we simulated phase transformations with two additional phase morphologies. We show the simulated and reconstruction phase morphologies in Figure 5. In Figure 5 a, the new phase (large-unit-cell, purple) nucleates from both sides of the nanoparticle and grows into the particle at the expense of the small-unit-cell (blue) structural phase. The input strain field (top) and the strain field retrieved by our algorithm from the simulated noisy X-ray data are again remarkably consistent. Apart from the few minor differences in the reconstructed shape, the reconstruction of the phase morphology is superb. A phase transformation following a core-shell structure (Figure 5 b) displays a similar quality. Therefore, we conclude that our algorithm is capable of reconstructing a variety of phase morphologies, provided the extent of the structural phase exceeds the spatial resolution of the method. The simulated X-ray flux distributes $10^6$ photons in the reciprocal space, which is accessible at current synchrotron sources with sub-micron crystalline particles.

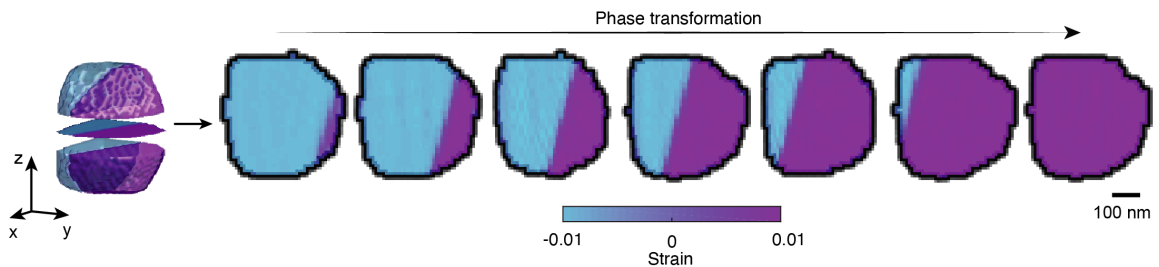

**Figure 4.** Results of the improved algorithm (see Figure 3) applied on the data shown in Figure 2b. Compare with Figure 2 a, which was the input for the simulation of the diffraction data in Figure 2b.

In summary, a combination of multiple strategies for improving phase retrieval in BCDI – the simultaneous reconstruction of several *in-situ* points, absolute amplitude smoothing, and parallel running of multiple reconstructions – enables the succesful reconstruction of 3D reciprocal space data for a nanoparticle with a strain magnitude of 1%. We have tested the algorithm on different phase morphologies: core-shell structure and different boundary nucleation morphologies. The algorithm reconstructs the structure successfully on all morphologies we tested. We find that for lower values of strain, turning off amplitude smoothing results in better reconstructions, which can be explained by the smoothing counteracting the shrinkwrap process of support tightening. Our method expands the capabilities of BCDI to the highly strained and phase-coexistent particles, and we expect that it will find use in *in-situ* and operando studies of intercalation materials and other nanomaterials exhibiting large strains.

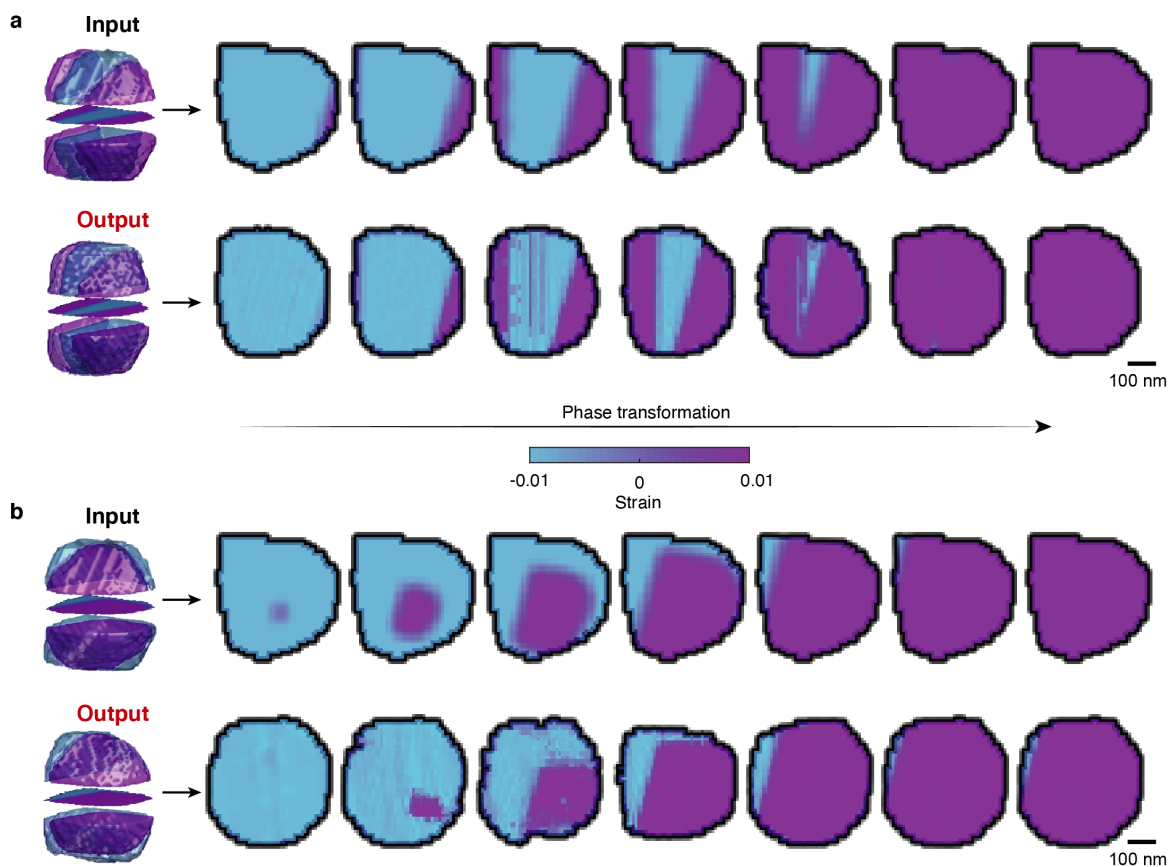

**Figure 5**: Simulated dynamic phase morophology (top) and phase retrieval result (bottom) from the simulated diffraction patterns (similar to Figure 2 and 4). (a) shows a striped domain configuration, (b) shows a core shell configuration. two different phase morphologies of phase transformations.

**Acknowledgements:**

This work was supported by the CCMR with funding from the NSF MRSEC program (DMR-1719875). We also acknowledge discussions with Lena Kourkoutis and Veit Elser from Cornell University.